\documentstyle[12pt]{article}
\newcommand{\ag}{\mbox{I \hspace{-0.82em} H}}
\def\bbox{{\,\lower0.9pt\vbox{\hrule \hbox{\vrule height 0.2 cm

\hskip 0.2 cm

\vrule  height 0.2 cm}\hrule}\,}}
\newcommand{\R}{\mbox{I \hspace{-0.82em} R}}
\def\bbox{{\,\lower0.9pt\vbox{\hrule \hbox{\vrule height 0.2 cm

\hskip 0.2 cm

\vrule  height 0.2 cm}\hrule}\,}}
\begin{document}
\setlength{\unitlength}{1mm}
\title{{\hfill {\small Alberta-Thy 20-97} } \vspace*{2cm} \\
Euclidean and Canonical Formulations of
Statistical Mechanics in
the Presence of Killing Horizons}
\author{\\
Dmitri V. Fursaev \date{}}
\maketitle
\noindent  {
{ \em
Theoretical Physics Institute, Department of
Physics, \ University of
Alberta \\ Edmonton, Canada T6G 2J1 and }
\\{\em Joint Institute for
Nuclear Research, Bogoliubov Laboratory of Theoretical
Physics, \\
141 980 Dubna, Russia}
\\
\\
e-mail: dfursaev@phys.ualberta.ca
}
\bigskip

\begin{abstract}
The relation between the covariant Euclidean
free-energy $F^E$ and the canonical
statistical-mechanical free energy $F^C$
in the presence of the Killing horizons is studied.
$F^E$ is determined by the
covariant Euclidean effective action.
The definition of $F^C$ is related to the
Hamiltonian which
is the generator of the evolution
along the Killing time.
At arbitrary temperatures $F^E$ acquires
additional ultraviolet divergences because of conical
singularities.
The divergences of $F^C$ are different
and occur since
the density ${dn \over d\omega}$ of the energy levels
of the system blows
up near the horizon in an infrared way.
We show that there are regularizations
that make it possible to remove the infrared cutoff
in ${dn \over d\omega}$.
After that the divergences of $F^C$ become
identical to the divergences of $F^E$.
The latter property turns out to be
crucial  to reconcile the covariant
Euclidean and the canonical
formulations of the theory.
The method we use is new and is based on a
relation between ${dn \over d\omega}$ and heat
kernels on hyperbolic-like
spaces.
Our analysis includes
spin 0 and spin 1/2 fields on arbitrary backgrounds.
For these fields the divergences of ${dn \over d\omega}$,
$F^C$ and $F^E$ are presented in the most complete form.
\end{abstract}

\bigskip

\baselineskip=.6cm

\newpage

\noindent
\section{Introduction}

There are two approaches
how
to describe quantum thermal effects
in the gravitational field.
The approach by Gibbons and Hawking
\cite{GiHa:76},\cite{Hawk:79}
defines the partition function
of the system as an Euclidean path integral.
It enables one to express the free energy
$F^E[g,\beta]$  in terms of the
effective action
$W[g,\beta]$, as $F^E[g,\beta]=\beta^{-1}W[g,\beta]$.
Functionals $W[g,\beta]$ are given
on Euclidean manifolds ${\cal M}_\beta$ with the period
$\beta$ in the Euclidean time $\tau$.
$\beta$ is considered as the inverse temperature.
The fields are assumed to be
periodic or antiperiodic in $\tau$,
depending on their statistics.
The Gibbons-Hawking approach is a straightforward
generalization of the
finite-temperature theory
in the Minkowsky space-time. Its advantage is that it
is manifestly
covariant in the Euclidean sector and
enables one to consider
the gravitational field on the equal footing with
matter fields.
This approach is especially convenient
for the thermodynamics
of black holes
\cite{GiHa:76}-\cite{BBWY:90}, it reproduces
the entropy, temperature
and other characteristics of a black hole
in the semiclassical approximation.

When the space-time is static
statistical-mechanical quantities can be also
described in a canonical  way. The canonical partition
function has the form
\begin{equation}\label{1.2}
Z^C=\mbox{Tr}~e^{-\beta \hat{\cal H}}~~~,
\end{equation}
where  the Hamiltonian $\hat{\cal H}$ is the generator
of the time evolution of the system.
$Z^C$ is well defined when
$\hat{\cal H}$ is a normally ordered
operator \cite{Allen:86}.
In this case the free-energy
takes the form
\begin{equation}\label{1.3}
F^C[g,\beta]=-\beta^{-1}\ln Z=
\eta\beta^{-1}\int_0^{\infty}d\omega {dn(\omega)
\over d\omega}\ln{(1-\eta e^{-\beta\omega})}~~~,
\end{equation}
where $\eta=+1$ for Bose fields and $\eta=-1$ for
Fermi fields.
${dn(\omega) \over d\omega}$ is the density
of eigen-values $\omega$
of {\it quantum-mechanical} Hamiltonians
of the fields
\cite{Allen:86}-\cite{DoSc:89}.
The advantage of definition
(\ref{1.3}) is that it is given in accordance with
the unitarity evolution of the system.
However, as distinct
from the Gibbons-Hawking approach, it is not
manifestly covariant.

There is no special
terminology\footnote{Allen \cite{Allen:86}
called $\exp(-\beta F^E)$ and
$\exp(-\beta F^C)$ the
"quantum" and "thermodynamic" partition functions.
These names are not very suitable
at least because
the both partition functions are
essentially quantum.}
to distinguish $F^E$
and $F^C$.
In this paper we call $F^E$ and
$F^C$ the covariant Euclidean and the canonical
free energies, respectively.
The corresponding formulations of the
finite-temperature theory will be called the
covariant Euclidean and the canonical formulations.
Sometime we will also say "Euclidean" instead of
"covariant Euclidean", for
simplicity.

For static space-times without horizons
comparing $F^E$ and $F^C$
shows \cite{Allen:86} that these functionals
differ only by the vacuum energy which
is not included in $F^C$.
Thus, in this case the Euclidean and canonical
formulations are, in fact, equivalent.

In space-times with Killing horizons
the quantum theory
has a number of specific properties.
On one hand, in the Euclidean formulation there
is a distinguished
value of the period $\beta$, corresponding
to the Hawking
temperature, for which ${\cal M}_\beta$ is a regular
space\footnote{This is the property of nonextremal
black hole backgrounds. The
extremal black holes will not be
considered here.}.
At other values of $\beta$ the space ${\cal M}_\beta$
has conical singularities
which result in additional ultraviolet divergences
\cite{Fursaev:94a}-\cite{Fursaev:95}.
On  another hand, the canonical formulation runs into
difficulties because the time evolution is not defined
at the bifurcation surface of the Killing horizons.
The density ${dn \over d\omega}$ of the energy levels
in Eq. (\ref{1.3}) blows up near the horizon
\cite{Hooft:85}-\cite{BCZ:96}
at any temperature in an infrared way.

As a result, in the presence of horizons the Euclidean
and the canonical free energies
look different and finding
the relation between them becomes a problem.
This problem has not
been analysed before and
our aim is to investigate it
for the case of scalar and spinor fields in some details.
Comparison of the covariant Euclidean and the
canonical formulations is important for
different reasons. The main of them is
statistical-mechanical
interpretation of black hole thermodynamics
which can be naturally defined in the framework of
the Gibbons-Hawking approach
\cite{GiHa:76}-\cite{BBWY:90}.

\bigskip

The paper is organized as follows.
In Section 2 we describe the Euclidean and
the canonical formulations of statistical
mechanics of scalar and
Dirac fields on static backgrounds.
The ultraviolet divergences appearing in $F^E$
because of conical singularities are
given in Section 3 in the most complete form.
We use the dimensional
and Pauli-Villars regularization procedures.
In Section 4
we develop a method how to find
the divergences of the density of levels
${d n \over d\omega}$.
We first apply this method to study
${d n \over d\omega}$ in the presence of
the spatial cutoff
near the horizon.
Then we show, in Section 5,  that
in dimensional and Pauli-Villars regularizations
the spatial cutoff can be removed and
one can define ${d n \over d\omega}$ on the complete
space. In these regularizations the divergences
of ${d n \over d\omega}$ have the ultraviolet
character. It means that the corresponding divergences
of the covariant free energy $F^C$ coincide exactly
with the divergences of the Euclidean free
energy $F^E$.
In Section 6 we discuss a hypothesis that
(for spinors and
scalars) the entire bare
functionals $F^C$ and $F^E$ must coincide.
We illustrate it with some examples.
Technical
details are given in Appendixes.
In Appendix A we remind the reader how
to relate the canonical free energy on ultrastatic
spaces
to the effective action. Appendix B
is devoted to the calculation of the
spinor heat coefficients
on conical singularities, some of the coefficients
represent the new result.

\section{Definitions and basic relations}
\setcounter{equation}0

Let us consider scalar fields $\phi$
described by the Klein-Gordon equation
and spinor fields
$\psi$ described by the Dirac equation,
\begin{equation}\label{2.1}
(-\nabla^\mu\nabla_\mu+\xi R +m^2)\phi=0~~~,
{}~~~
(\gamma^\mu\nabla_\mu+m)\psi=0~~~,
\end{equation}
where $R$ is the scalar curvature and
$\nabla_\mu$ are the covariant
derivatives,
defined according with the spin of the
fields\footnote{We define the spinor
derivative as $\nabla_\mu=\partial_\mu+\Gamma_\mu$,
where
$\Gamma_\mu=
\frac 18[\gamma^\lambda,\gamma^\rho]~V^i_\rho
\nabla_\mu V_{i\lambda}$ is the connection
and $V^i_\nu$ are the tetrades.}.
The Dirac $\gamma$-matrices
$\gamma^\mu=(\gamma^0,\gamma^a)$ obey
the standard commutation relations
$\{\gamma^\mu,\gamma^\nu\}=2g^{\mu\nu}$,
$\gamma^0$ is the anti-Hermitean matrix.
It is supposed that the space-time is static
\begin{equation}\label{1.4}
ds^2=g_{00}dt^2+g_{ab}dx^a dx^b~~~,~~~a,b=1,2,3~~~.
\end{equation}
The component $g_{00}$ is a nonpositive
function and
$g_{00}=-1$ at spatial infinity.
The temperature measured at
infinity is $\beta^{-1}$.
On space (\ref{1.4})
equations (\ref{2.1}) can be rewritten in the form
\begin{equation}\label{2.2}
-g^{00}(\partial_t^2+H_s^2)\phi=0~~~,
{}~~~
-i\gamma^0(i\partial_t+H_d)\psi=0~~~,
\end{equation}
\begin{equation}\label{2.3}
H_s^2=
|g_{00}|(-\nabla^a\nabla_a -
w^a\nabla_a+m^2+\xi R)~~~,
\end{equation}
 \begin{equation}\label{2.4}
H_d=i\gamma_0\left(\gamma^a
(\nabla_a+\frac 12 w_a)+m\right)~~~.
\end{equation}
Here $\nabla_a$
is the covariant derivative computed
with the help of the metric $g_{ab}$
of the three-dimensional surface of constant time
$t=const$.  We denote this
surface ${\cal B}$.
Index $a$ is up and down with the help of
$g_{ab}$. The vector $w_a=\frac 12
\nabla_a\ln |g_{00}|$ is the vector of
acceleration.
The operators $H_s$ and $H_d$ are called the
one-particle
Hamiltonians because their eigen-values
coincide with
the frequencies of one-particle excitations
\footnote{
It is easy to check that $H_s^2$ and $H_d$ are
Hermitean operators
with respect to the following inner products
$
(\phi',\phi)=\int_{\cal B}
\sqrt{^{(3)}g|g_{00}|^{-1}}
{}~d^3x~
(\phi')^{*}\phi
$,
$
(\psi',\psi)=\int_{\cal B}\sqrt{^{(3)}g}
{}~d^3x~
(\psi')^{+}\psi
$
where $^{(3)}g=\det g_{ab}$ and
$(\psi')^{+}$ denotes Hermitean conjugation.}.
To calculate the canonical free energies
$F^C_i$  ($i=s,d$) with the help of
Eq. (\ref{1.3}) one has to know the densities
${dn_i \over d\omega}$ of the energy levels of
$H_s$ and $H_d$.

Let us define now the Euclidean free energies
for the fields described by Eq. (\ref{2.1}).
To this aim we
consider the Euclidean manifold ${\cal M}_\beta$
which is the Euclidean section of the
Lorentzian geometry (\ref{1.4})
\begin{equation}\label{2.5}
ds^2=g_{\tau\tau}d\tau^2 +
g_{ab}dx^a dx^b~~,~~0\leq\tau\leq \beta~~,
\end{equation}
where $g_{\tau\tau}=|g_{00}|$.
The {\it Euclidean} effective actions
for the fields $\phi$ and $\psi$ are
\begin{equation}\label{2.7}
W_s[g,\beta]=
{1\over 2} \log\det L_s~~~,~~~W_d[g,\beta]=
-\log\det L_d~~~,
\end{equation}
\begin{equation}\label{2.6}
L_s=-\nabla^\mu\nabla_\mu+\xi R +m^2~~~,~~~
L_d=\gamma_5(\gamma^\mu\nabla_\mu+m)~~~.
\end{equation}
It is assumed that $W_i$ are regularized functionals.
The operators $L_s$ act on scalar fields on
${\cal M}_\beta$ which are periodic
in $\tau$,
$L_d$ act on spinors which change the sign
when $\tau$ is increased by $\beta$.
The Euclidean matrix $\gamma_\tau$ is
$i\gamma_0$, the matrix $\gamma_5$
anticommutes with the other $\gamma$'s
and it is "normalized" as $\gamma_5^2=1$.
Both operators (\ref{2.6})
are Hermitean with respect to the standard
inner product
$(\varphi,\varphi')=
\int \varphi^{+}\varphi'\sqrt{g}d^4x$.
According to Eq. (\ref{1.3}), the canonical
free energy vanishes at zero temperature.
It is convenient to define the Euclidean free energy
so that it have the same property, i.e., as
\begin{equation}\label{2.8}
F^E_i[g,\beta]=\beta^{-1}W_i[g,\beta]-E^0_i[g]~~~,
\end{equation}
\begin{equation}\label{2.9}
E^0_i[g]=\lim_{\beta\rightarrow\infty}\left(\beta^{-1}
W_i[g,\beta]\right)~~~.
\end{equation}
The quantities $E^0_i[g]$ have the meaning of the vacuum
energy.
Note that
$F^E_i[g,\beta]$ and $E^0_i[g,\beta]$
are covariant functionals of the metric, because
the Euclidean actions $W_i[g,\beta]$ are covariant at
any values of $\beta$.

\bigskip
Our aim is to find the relation between
$F^E$ and $F^C$.
The important property of $F^C$ is that it can be
represented in the form similar to Eq. (\ref{2.8})
\begin{equation}\label{2.10}
F^C_i[g,\beta]=\beta^{-1}\bar{W}_i[g,\beta]-
\bar{E}^0_i[g]~~~,
\end{equation}
\begin{equation}\label{2.11}
\bar{E}^0_i[g]=
\lim_{\beta\rightarrow\infty}\left(\beta^{-1}
\bar{W}_i[g,\beta]\right)~~~.
\end{equation}
The functionals $\bar{W}_i[g,\beta]$ are the following
effective actions
\begin{equation}\label{2.12}
\bar{W}_s[g,\beta]=
{1\over 2} \log\det \bar{L}_s~~~,
{}~~~\bar{W}_d[g,\beta]=
-\log\det \bar{L}_d~~~.
\end{equation}
The operators $\bar{L}_i$ are related to
$L_i$, Eq. (\ref{2.6}), by the conformal transformations
\begin{equation}\label{2.13}
\bar{L}_s=
e^{-3\sigma}L_s~e^{\sigma}~~~,~~~
\bar{L}_d=e^{-\frac 52 \sigma}L_d~
e^{\frac 32 \sigma}~~~,
\end{equation}
where $e^{-2\sigma}=g_{\tau\tau}$.
Representation (\ref{2.10}) is well known
\cite{Allen:86}-\cite{Dowker:84} but
for the sake of completeness we give its derivation
in Appendix A. The key point is that $\bar{L}_i$
are expressed in terms of the one-particle Hamiltonians
\begin{equation}\label{2.14}
\bar{L}_s=\bar{H}_s^2-\partial^2_\tau~~~,
\end{equation}
\begin{equation}\label{2.15}
\bar{L}_d=\gamma_5\bar{\gamma}_\tau
(\bar{H}_d+\partial_\tau)~~~,~~~~
\bar{L}_d^2=\bar{H}_d^2-\partial^2_\tau~~~,
\end{equation}
\begin{equation}\label{2.16}
\bar{H}_s^2=e^{-\sigma}H_s^2e^{\sigma}~~~,~~~
\bar{H}_d=e^{-\frac 32 \sigma}~
H_d ~e^{\frac 32 \sigma}~~~,
\end{equation}
where $\bar{\gamma}_\tau=e^\sigma\gamma_\tau$.
Note that according to Eqs. (\ref{2.16}) the spectra
of the operators $\bar{H}_i$ and $H_i$, Eqs.
(\ref{2.3}), (\ref{2.4}),
coincide, which means that
$\bar{H}_i$ are simply an another
representation of $H_i$.
The operators $\bar{L}_i$ act on the fields on
the ultrastatic space $\bar{\cal M}_\beta$ with the
metric
\begin {equation}\label{2.17}
d\bar{s}^2=d\tau^2 + \bar{g}_{ab} dx^a dx^b~~~,
{}~~~0 \leq \tau \leq \beta~~~
\end {equation}
which is conformally related to metric (\ref{2.5}),
$\bar{g}_{\mu\nu}=g_{\mu\nu}/|g_{00}|$.
One can show that
\begin{equation}\label{2.18}
\bar{H}_d=i\bar{\gamma_0}(\bar{\gamma}^a
\bar{\nabla}_a+e^{-\sigma}m )~~~,
\end{equation}
where $\{\bar{\gamma}^{\mu},\bar{\gamma}^{\nu}\}
=2\bar{g}^{\mu\nu}$, and
\begin{equation}\label{2.19}
\bar{H}_i^2=-\bar{\nabla}^a
\bar{\nabla}_a+e^{-2\sigma}m^2+V_i~~~.
\end{equation}
The derivatives $\bar{\nabla}_\alpha$ are defined
with the help of the metric $\bar{g}_{ab}$ of
the surface $\tau=const$ in ultrastatic space
(\ref{2.17}).
This surface is conformally related to ${\cal B}$
and we denote it $\bar{\cal B}$.
The
"potential terms" $V_i$ are
\begin{equation}\label{2.20}
V_s=\xi\bar{R}+
e^{-2\sigma}(1-6\xi)
\left(\nabla^\mu w_\mu-w^\mu w_\mu\right)
{}~~~,
\end{equation}
\begin{equation}\label{2.21}
V_d=\frac 14 \bar{R}+e^{-2\sigma}m
\gamma^\mu w_\mu~~~.
\end{equation}
Here $w_\mu=(0,w_a)$ is the four vector of
acceleration and $\bar{R}$ is the
curvature of the ultrastatic
background
\begin{equation}\label{2.22}
\bar{R}=e^{-2\sigma}\left(R+
6(\nabla^\mu w_\mu-w^\mu w_\mu)\right)
{}~~~.
\end{equation}
The formulation
of statistical mechanics in terms of
the theory on ultrastatic spaces
was developed by Dowker and
Kennedy \cite{DoKe:78}
and by Dowker and Schofield
\cite{DoSc:88},\cite{DoSc:89}.
Relations (\ref{2.18})-(\ref{2.22})
coincide with
those used in Refs.\cite{DoSc:88},\cite{DoSc:89}.

The actions $W_i$ and $\bar{W}_i$
are determined by the conformally related
wave operators, see Eqs. (\ref{2.13}).
In static space-times without horizons the
renormalized
functionals $W_i$ and $\bar{W}_i$ differ
by the anomalous terms computed in
\cite{DoSc:88},\cite{DoSc:89}. These terms,
however, are
proportional to $\beta$, so they
result in the difference between  vacuum energies
(\ref{2.9}) and (\ref{2.11}). The Euclidean
and the canonical free energies in
this case coincide.
When there is a horizon the
conformal transformation
to the ultrastatic metric becomes singular
and this case
requires a special analysis.

\section{Covariant Euclidean formulation:
conical singularities
and divergences}
\setcounter{equation}0

To begin with we describe the class of space-times
with Killing horizons which will be discussed here.
We assume that the metric (\ref{1.4})
near the bifurcation surface
$\Sigma$ has the following form
\begin{equation}\label{4.8}
ds^2=g_{00}(\theta,\rho)dt^2+d\rho^2+
\gamma_{pq}(\theta,\rho)d\theta^pd\theta^q~~~,
{}~~~p,q=1,2~~~.
\end{equation}
In this representation the location
of $\Sigma$ is determined by
the equation $\rho=0$,  $\theta^p$ are the
coordinates on this surface. We also
assume that
near $\rho=0$ the components of the metric are
decomposed as
\begin{equation}\label{4.9}
g_{00}(\theta,\rho)=-
\kappa^2\rho^2\left(1-\frac 16 \rho^2 R_{ijij}(\theta)+
O(\rho^4)\right)~~~,
\end{equation}
\begin{equation}\label{4.10}
\gamma_{pq}(\theta,\rho)=\tilde{\gamma}_{pq}(\theta)+
\rho^2h_{pq}(\theta)
+O(\rho^4)~~~,
\end{equation}
where $\tilde{\gamma}_{pq}$ is the metric tensor
on $\Sigma$. The constant $\kappa$ is called
the surface gravity. Metrics which
obey the properties
(\ref{4.8})-(\ref{4.10}) correspond to
static nonextremal
black holes. It can be shown that the
quantities
$R_{ijij}$, $R_{ii}$ are the
projections of the Riemann and Ricci
tensors calculated
on $\Sigma$ on the directions normal
to this surface.
Namely,
\begin{equation}\label{3.3}
R_{ii}=
R_{\mu\nu}n^{\mu}_i n^{\nu}_i~~~,~~~
R_{ijij}=R_{\mu\lambda\nu\rho}n^{\mu}_i n^{\lambda}_j
n^{\nu}_i n^{\rho}_j~~~.
\end{equation}
Here $n^\mu_i$ are two unit orthonormal vectors
orthogonal to $\Sigma$ and the summation over the
indexes $i,j=1,2$
is assumed. It can be also shown that
\begin{equation}\label{4.11}
h_{pq}\tilde{\gamma}^{pq}=
\frac 12 (R_{ijij}-R_{ii})~~~.
\end{equation}
Near $\Sigma$ the Euclidean section
${\cal M}_\beta$ of space-time (\ref{4.8})
looks as
\begin{equation}\label{3.1}
ds^2\simeq \kappa^2\rho^2 dt^2+d\rho^2 +
\tilde{\gamma}_{pq}d\theta^p
d\theta^q~~~,~~~
0\leq\tau\leq \beta~~~.
\end{equation}
This space is regular when
$\beta^{-1}=\beta^{-1}_H={\kappa \over 2\pi}$, where
the constant $\beta^{-1}_H$ is called the Hawking
temperature. For arbitrary
$\beta$ there are conical
singularities and ${\cal M}_\beta$ looks as
${\cal C}_\beta\times \Sigma$, where ${\cal C}_\beta$
is a cone.

As a result of conical singularities,
the Euclidean free energy $F^E$ is divergent
even after subtracting from it
the vacuum energy.
The divergent part $F^E_{\mbox{div}}$ of $F^E$ can be
calculated with the help of different regularizations.
We begin with the dimensional regularization
and consider $D$-dimensional space-time.
It is assumed that when going to arbitrary
dimensions the
background space holds its Killing structure
and equations (\ref{4.8})-(\ref{4.10}) do not change.
The difference now is that the tensors
$R_{\mu\nu\lambda\rho}$, $R_{\mu\nu}$
are calculated in $D$ dimensions and $\Sigma$ is
a $(D-2)$
dimensional surface.
The divergent part of the Euclidean free energy is
\begin{equation}\label{3.2}
F^E_{\mbox{div}}[g,\beta,D]=-\eta
{\Gamma\left(1-\frac D2\right) \over (4\pi)^{D/2}}
{\pi^2 m^{D-4} \over 3\kappa \beta^2}
\int_{\Sigma}
\left[f_1 m^2- \left(p_1
{4\pi^2 \over \kappa^2\beta^2} P +
p_2 R+p_3 R_{ii}\right)
\right]~~~,
\end{equation}
where $D$ is considered as a complex parameter
and the integral is taken over the bifurcation surface
$\Sigma$, $\int_{\Sigma}\equiv
\int_{\Sigma}\sqrt{\tilde{\gamma}}d^{D-2}\theta$.
We put $P=2R_{ijij}-R_{ii}$ and introduce the
constants $f_1$ and $p_k$ which are
listed in Table 1:

\renewcommand{\baselinestretch}{2}
\begin{center}
{\small

\begin{tabular}{|c||c|c|c|c|}
\hline
$\mbox{spin}$  & $f_1$
& $p_1$   & $p_2$   & $p_3$  \\
\hline
\hline
$0$ &  $1$   &
${1 \over 60}$ & $\frac 16-\xi$  & $0$ \\
\hline
\hline
$\frac 12$ & $-\frac 12 r_d$ &
$-{7 \over 480}r_d$ & ${1 \over 24}r_d$  &
$-{1 \over 16}r_d$ \\
\hline
\end{tabular}}
\end{center}
\begin{center}
Table 1.
\end{center}

\renewcommand{\baselinestretch}{1}
\noindent
Here $r_d$ is the dimensionality of the spinor
representation, $r_d=4$ for
Dirac spinors and $r_d=2$
for massless Weyl spinors.
As follows from Eq. (\ref{3.2}), in
the dimensional regularization $F^E$
has a simple pole at $D=4$.
The dimensional regularization
reproduces the divergences of the logarithmical type
only.
For this reason it is also worth studying
$F^E_{\mbox{div}}$ in Pauli-Villars
regularization
which usually gives all divergent terms.
The Pauli-Villars method is based on introduction of
several, say, $5$
additional fields. $2$ fields with masses $M_k$
have the same statistics as the original
field, while other $3$
ones with masses $M_r'$ have the wrong statistics,
i.e., they are fermions for scalars
and bosons for spinors. The latter fields
give contribution to
$F^E$ with the sign opposite to
that of physical fields. To eliminate the
divergences  two restrictions are imposed
\begin{equation}\label{3.4}
m^{p}+\sum_k M_k^{p}-\sum_r(M'_r)^{p}=0~~~,
{}~~~p=2,4~~~.
\end{equation}
These equations can be resolved by choosing
\cite{DLM:95}
$M_{1,2}=\sqrt{3\mu^2+m^2}$,
$M'_{1,2}=\sqrt{\mu^2+m^2}$,
$M'_3=\sqrt{4\mu^2+m^2}$.
The divergences
in Pauli-Villars regularization can be obtained
from Eq. (\ref{3.2}). By adding contributions
of the regulator fields with the sign corresponding to
their
statistics and taking the limit $D\rightarrow 4$,
which is finite due to restriction (\ref{3.4}) with
$p=2$, one finds
\begin{equation}\label{3.5}
F^E_{\mbox{div}}[g,\beta,\mu]=-
{\eta \over 48\kappa \beta^2}
\int_{\Sigma}
\left[b f_1 + a \left(p_1
{4\pi^2 \over \kappa^2\beta^2} P +
p_2 R+p_3 R_{ii}\right)
\right]~~~,
\end{equation}
\begin{equation}\label{3.5b}
a=a(m,\mu)=-
\ln m^2-\sum_k \ln M_k^2+
\sum_r \ln (M'_r)^2~~~,
\end{equation}
\begin{equation}\label{3.5c}
b=b(m,\mu)=
m^2\ln m^2+\sum_k  M^2_k\ln M_k^2-
\sum_r (M'_r)^2\ln (M'_r)^2~~~.
\end{equation}
The parameter $\mu^2$ plays the role of
the ultraviolet cutoff. The
regularization is removed
when $\mu\rightarrow \infty$. In this limit
$a\sim \ln (\mu^2/m^2)$ and $b\sim \mu^2$ ($a,b>0$).
Thus, in general, $F^E_{\mbox{div}}$
includes both logarithmic and
quadratic divergences.

\bigskip

The derivation of Eq. (\ref{3.2}) is standard.
In the dimensional regularization the
Schwinger-DeWitt proper-time
representation \cite{DeWitt:65} gives
the divergent part of the effective action in the
form
\begin{equation}\label{3.6}
W^E_{\mbox{div}}[g,\beta]=
-{\eta \over 2}
\int_{0}^{\infty}{ds \over s}
e^{-m^2s}{1 \over (4\pi s)^{D/2}}
\left(B_0+s B_1+s^2 B_2\right)~~~.
\end{equation}
Here $B_k$ are the
Hadamard-Minackshisundaram-DeWitt-Seeley
(or heat) coefficients of the heat kernel
asymptotic
expansion\footnote{We do not take
into account the boundaries of
space-time.}
\begin{equation}\label{3.7}
\mbox{Tr}~e^{-s\triangle}\approx
{1 \over (4\pi s)^{D/2}}
\left(B_0+sB_1+s^2 B_2+...\right)~~~.
\end{equation}
The Laplacians $\triangle$ look as
$\triangle=-\nabla^\mu\nabla_\mu + X$, where
$\nabla_\mu$ is the covariant derivative
defined according with the spin;
$X=(1/6-\xi) R$ for scalars,
and $X=\frac 14 R$ for spinors.  The relation
between $\triangle$'s and the operators
$L_s$ and $L_d$,
Eqs. (\ref{2.6}), is
\begin{equation}\label{3.8}
L_s=\triangle_s+m^2~~~,
{}~~~L_d^2=\triangle_d+m^2~~~.
\end{equation}
On singular spaces ${\cal M}_\beta$ the coefficients
$B_k$ can be represented
as the sum of two terms
\begin{equation}\label{3.9}
B_k=A_k+A_{\beta,k}~~~.
\end{equation}
$A_k$ have the form of the standard coefficients
defined on the regular domain of ${\cal M}_\beta$,
$A_{\beta,k}$, $k\geq 1$, are functionals
on $\Sigma$ which appear because of
conical singularities. For scalars and
spinors the first two coefficients
have the form
\begin{equation}\label{3.10}
A_{\beta,1}=f_1
{\pi \over 3\gamma}(\gamma^2-1)
{\cal A}~~~,
\end{equation}
\begin{equation}\label{3.11}
A_{\beta,2}={\pi \over 3\gamma}\int_{\Sigma}
\left[(\gamma^4-1)p_1 P+(\gamma^2-1)(p_2 R+p_3 R_{ii})
\right]~~~,
\end{equation}
where $\gamma={\beta_H \over \beta}$ and the numbers
$f_1$ and $p_k$ are given in Table 1.
Expression (\ref{3.5}) for $F^E_{\mbox{div}}[g,\beta]$
follows from Eqs. (\ref{2.8}), (\ref{2.9}) and
(\ref{3.6}),(\ref{3.9})-(\ref{3.11}).

For scalars the coefficient $A_{\beta,1}$
was found  by Cheeger \cite{Cheeger:83},
see also Refs. \cite{CKV:94},\cite{Fursaev:94a}.
The spinor coefficient $A_{\beta,1}$
follows from the results of Refs.
\cite{Kabat:95},\cite{FM:97}.
The scalar coefficient $A_{\beta,2}$
and the general structure
of the higher coefficients $A_{\beta,k}$
were analysed in Refs.
\cite{Fursaev:94b},\cite{Dowker:94a},\cite{Dowker:94b}.
The calculation of the spinor coefficient
$A_{\beta,2}$
is our new result. Its derivation is similar
to that of Ref. \cite{Fursaev:94b}
but has new features related to the spin. The reader
can find the details in Appendix B.

\section{Canonical formulation: infrared
divergences}
\setcounter{equation}0

Our aim now is to investigate the divergences
of the canonical free energy  and
compare them with the results
(\ref{3.2}),(\ref{3.5}) found in the Euclidean
formulation. We begin with remarks concerning
specific features of
quantum systems in the presence of horizons.
In this case the one-particle oscillators of
fields have a continuous spectrum
of frequencies
$0\leq \omega < \infty$. The eigenvalues of
the operators $\bar{H}_i$, Eqs.
(\ref{2.18}),(\ref{2.19}),
run down to $\omega=0$
even for massive fields, i.e., the usual
mass gap is absent. This property has the
simple explanation. $\bar{H}_i$ are
given on the
space $\bar{\cal B}$, which is the spatial
part of ultrastatic space (\ref{2.17}) related
to the original space (\ref{2.5}) by the
conformal transformation. On the
ultrastatic space the location
of the horizon is
mapped at infinity and $\bar{\cal B}$
turns out to be non-compact. At the same time,
the masses $m_i$ of the fields can be neglected
near the horizon because they are multiplied
by the vanishing factor $e^{-2\sigma}=|g_{00}|$.
Regarding the "potential" terms in $\bar{H}_i$,
see Eqs. (\ref{2.20}),(\ref{2.21}),
they are constant and negative on the horizon,
$V_s=-\kappa^2$, $V_d= -\frac 32\kappa^2$,
and look as tachionic masses
\cite{CVZ:95},\cite{BCZ:96}.
As a result,
the densities ${dn_i \over d\omega}$ of
the eigen-values
of $\bar{H}_i$ blow up
near the horizon in an infrared way.

\bigskip

To investigate this divergence
the following method
can be used.
By the definition, the trace of the heat
kernel of the operator $\bar{H}_i^2$
is
\begin{equation}\label{4.1}
\mbox{Tr}~e^{-\bar{H}_i^2t}=\int_{0}^{\infty}
d\omega {dn_i(\omega) \over d\omega}
e^{-\omega^2t}~~~.
\end{equation}
By making use of the inverse Laplace
transform one
obtains
\begin{equation}\label{4.2}
{dn_i(\omega) \over d\omega}=
{\omega \over \pi}\int_{-\infty}^{+\infty}
d\alpha~ e^{i\alpha\omega^2}\mbox{Tr}~
e^{-i\alpha\bar{H}_i^2}~~~.
\end{equation}
The diagonal matrix elements
$\langle x|\exp(-\bar{H}_i^2t)|x \rangle
\equiv \left[\exp(-\bar{H}_i^2t)
\right]_{\mbox{diag}}$
are well defined but
the traces involve the
integration over the non-compact
space $\bar{\cal B}$
and diverge. The key observation is that the
divergent parts of ${dn_i(\omega) \over d\omega}$
can be found
by making use of the asymptotic
properties of the traces
at small $t$, which are very well known.

Let us begin with massless
fields in the Rindler space
\begin{equation}\label{4.3a}
ds^2=-\kappa^2\rho^2 dt^2 +d\rho^2+dx^2+dy^2
{}~~~,~~
-\infty <x,y<\infty~~,~~\rho > 0~~.
\end{equation}
The Rindler horizon is the plane $\R^{2}$,
$\rho$ is the proper distance to
the horizon.
The Rindler space can be considered as an
approximation to
the black hole geometry near the horizon.
The metric on the corresponding space
$\bar{\cal B}$ is
\begin{equation}\label{4.3}
dl^2=\kappa^{-2}\rho^{-2}(d\rho^2+dx^2+dy^2)~~~.
\end{equation}
As can be shown, $\bar{\cal B}$ coincides with the
hyperbolic manifold $\ag^{3}$ having the
constant curvature
$\bar{R}=-6\kappa^2$ \cite{BCVZ:96}.
A review of the heat kernels of Laplace operators on
such spaces can be found in Refs.
\cite{BCVZ:96},\cite{Camporesi:90}.
It is remarkable that for the massless fields the
heat kernels are known explicitly and their
diagonal elements read\footnote{The expression
for the scalar kernel is given in Ref. \cite{BCVZ:96},
the spinor kernel follows from
the $\zeta$ function which is
also given there.}
\begin{equation}\label{4.4}
\left[e^{-t\bar{H}_s^2}\right]_{\mbox{diag}}
={1 \over (4\pi t)^{3/2}}~~~,
{}~~~
\left[ e^{-t\bar{H}_d^2}\right]_{\mbox{diag}}
={r_d \over (4\pi t)^{3/2}}
\left[1+\frac 12 \kappa^2 t\right]~~~.
\end{equation}
The number $r_d$ appears in the spinor case
after tracing over the
spinor indexes. Obviously, the
traces of these operators diverge at $\rho=0$.
Let us restrict the integration in the traces
by values
$\rho\geq \epsilon$ where $\epsilon$
is a proper distance
to the horizon. Such a method is called
the volume cutoff.
With the help of Laplace transform (\ref{4.2})
(see Ref. \cite{Bateman:54})
one easily obtains from (\ref{4.4}) the divergences
of the densities of levels at
$\epsilon\rightarrow 0$
\begin{equation}\label{4.5}
\left[{dn_s(\omega,\epsilon) \over d\omega}\right]
_{\mbox{div}}
= {{\cal A} \over 4\pi^2 \kappa^3}
{\omega ^2 \over \epsilon^2}
{}~~~,~~~
\left[{dn_d(\omega,\epsilon) \over d\omega}\right]
_{\mbox{div}}=r_d
{{\cal A} \over 4\pi^2 \kappa^3}\left[
{\omega ^2 \over \epsilon^2}+
{\kappa^2 \over 4\epsilon^2}
\right]~~~
\end{equation}
where ${\cal A}$ formally stands for the
area of the horizon
\footnote{Since the Rindler horizon
is a plane, only the densities of levels per unit
area have a strict meaning.}.
For scalar fields our result agrees
with previous computations by the
WKB method, see, for instance Ref. \cite{Hooft:85}.
A generalization of (\ref{4.5}) to massive scalars
was explicitly
found in Refs. \cite{CVZ:95},\cite{BCZ:96}.

Let us consider how do Eqs. (\ref{4.5}) modify
when the geometry deviates from the Rindler form.
The spaces $\bar{\cal B}$ can be approximated
by $\ag^3$ only
in the limit $\rho\rightarrow 0$. For this
reason, the diagonal
elements of the heat kernels of
$\bar{H}_i^2$ on $\bar{\cal B}$ are represented
by the Taylor
series in $\rho^2$ converging at
$\rho\rightarrow 0$ to expressions
(\ref{4.4}). Presumably,
the coefficients
in these series should be the local
functions of the curvature and so to find
them
it is sufficient to use the asymptotic form of the
traces. One has
\begin{equation}\label{4.6}
\left[ e^{-\bar{H}_i^2t} \right]_{\mbox{diag}} \simeq
{1 \over (4\pi t)^{3/2}}\left(r_i+\bar{a}_{i,1}t+
\bar{a}_{i,2}t^2+...\right)~~~,
\end{equation}
where $r_s=1$ and $\bar{a}_{i,n}$ are the
diagonal elements of the corresponding
heat coefficients. As before the summation over the
spinor indexes in (\ref{4.6}) is assumed.
With the help of Eqs. (\ref{2.20})-(\ref{2.22})
one finds
\begin{equation}\label{4.7a}
\bar{a}_{s,1}=e^{-2\sigma}\left[\left(\frac 16-
\xi\right) R-m^2\right]~~~,
\end{equation}
\begin{equation}\label{4.7b}
\bar{a}_{d,1}=-e^{-2\sigma}r_d
\left[{1 \over 12} R
+\frac 12 (\nabla^\mu w_\mu -w^\mu w_\mu)
+m^2\right]~~~.
\end{equation}
These expressions can be decomposed in powers
of $\rho^2$ by making use of Eqs.
(\ref{4.8})-(\ref{4.11})
\begin{equation}\label{4.12}
\nabla^\mu w_\mu=-R^t_t=-\frac 12
R_{ii}+O(\rho^2)~~~,
\end{equation}
\begin{equation}\label{4.13}
w^\mu w_\mu={1 \over \rho^2}\left(
1-\frac 13 \rho^2 R_{ijij}+O(\rho^4)\right)~~~,
\end{equation}
\begin{equation}\label{4.14}
\bar{a}_{s,1}=\kappa^2\rho^2
\left[\left(\frac 16-\xi\right) R-m^2\right]
+O(\rho^4)~~~,
\end{equation}
\begin{equation}\label{4.15}
\bar{a}_{d,1}=r_d {\kappa^2 \over 2}
\left[1+\rho^2\left(\frac 12 R_{ii}-
\frac 12 R_{ijij}-\frac 16 R-2m^2\right)+
O(\rho^4)\right]~~~,
\end{equation}
where $R$ is evaluated at $\rho=0$.
For the massless fields in the Rindler
space  $\bar{a}_{s,1}=0$,
while $\bar{a}_{d,1}=r_d\frac 12
\kappa^2$, in agreement
with Eqs. (\ref{4.4}).
Decompositions (\ref{4.14}), (\ref{4.15})
are written explicitly
up to the terms of the order
$\rho^2$, other terms do not contribute to
the divergence of the traces.

Analogous decompositions can be found
for the heat coefficients
$\bar{a}_{i,n}$ with $n\geq 2$, however,
$\bar{a}_{i,n}$ vanish at small $\rho$
faster than $\rho^2$.
To see this let us consider the
operators $\bar{L}_s$, $\bar{L}_d^2$
which are given on the ultrastatic
space $\bar{\cal M}_\beta$, Eq. (\ref{2.17}).
$\bar{L}_s$, $\bar{L}_d^2$ are related to the
three dimensional
Hamiltonians $\bar{H}_i^2$ by Eqs.
(\ref{2.14}),(\ref{2.15}).  Because
${\bar{\cal M}}_\beta=
S^1\times\bar{\cal B}$
the diagonal parts of the heat
coefficients of $\bar{L}_s$, $\bar{L}_d^2$
coincide with $\bar{a}_{i,n}$. On the other hand,
$\bar{L}_s$, $\bar{L}_d^2$ are related by
conformal transformation (\ref{2.13})
to the covariant
Euclidean operators $L_s$, $L_d^2$.
In this case, according to Dowker
and Schofield \cite{DoSc:88},\cite{DoSc:89},
\begin{equation}\label{4.16}
\bar{a}_{i,2}=e^{-4\sigma}\left(a_{i,2}+\nabla_\mu
J^\mu_{i}\right)~~~,
\end{equation}
where $a_{i,2}$ are the heat coefficients
corresponding to $L_s$, $L_d^2$. The currents
$J^\mu_{i}$ have the
form\footnote{Note that
our signature of the metric is different from
that in Refs. \cite{DoSc:88},\cite{DoSc:89}.}
$$
J^\mu_s=-{1 \over 45}\left\{5\left[\frac 12
\nabla^\mu(w^2)+w^\mu(w^2)-w^\mu\nabla w\right]
-\frac 32\nabla^\mu\nabla w-2R^{\mu\nu}w_\nu
\right.
$$
\begin{equation}\label{4.17}
\left.
+\frac 32 w^\mu R
+15\left[\left(\xi-\frac 16\right)R+m^2\right]w^\mu
\right\}~~~,
\end{equation}
$$
J^\mu_d=-{r_d \over 45}\left\{-14\left[\frac 12
\nabla^\mu(w^2)+w^\mu(w^2)-w^\mu\nabla w\right]
+18\nabla^\mu\nabla w+22R^{\mu\nu}w_\nu
\right.
$$
\begin{equation}\label{4.18}
\left.
-(5R+30m^2)w^\mu
\right\}~~~,
\end{equation}
where $w^2=w^\nu w_\nu$ and $\nabla w=
\nabla^\nu w_\nu$.
The coefficients $a_{i,2}$
are regular at $\rho\rightarrow 0$.
With the help of Eqs. (\ref{4.12}),(\ref{4.13})
one can check that
$\nabla_\mu J^\mu_{i}$ are regular as well.
Therefore, as
follows from (\ref{4.16}),
$\bar{a}_{i,2}\sim e^{-4\sigma}\sim \rho^4$.
The other coefficients $\bar{a}_{i,n}$
with $n>2$
are determined from $\bar{a}_{i,2}$ by
recursion
relations \cite{DeWitt:65}
and so $\bar{a}_{i,n}$ vanish near
$\Sigma$ at least as
fast as $\bar{a}_{i,2}$.

Our conclusion is that
only
$\bar{a}_{i,1}$ contribute to the
divergences of the
traces.
{}From Eqs.
(\ref{4.2}),(\ref{4.6}),(\ref{4.14}),(\ref{4.15})
we obtain the
divergent part of the densities of levels
\begin{equation}\label{4.19}
\left[{dn_s(\omega,\epsilon) \over d\omega}\right]
_{\mbox{div}}
= {1 \over 4\pi^2 \kappa^3}\int_{\Sigma}
\left\{\omega ^2
\left({1 \over \epsilon^2}-\frac 14 P
\ln {\epsilon^2 \over l^2}\right)
-{\kappa^2 \over 2}
\ln {\epsilon^2 \over l^2}
\left[\left(\frac 16-\xi\right) R-m^2\right]\right\}
{}~~~,
\end{equation}
$$
\left[{dn_d(\omega,\epsilon) \over d\omega}\right]
_{\mbox{div}}=
{r_d \over 4\pi^2 \kappa^3}\int_{\Sigma}
\left\{
\omega ^2
\left({1 \over \epsilon^2}-\frac 14 P
\ln {\epsilon^2 \over l^2}\right)
+{\kappa^2 \over 4\epsilon^2}
\right.
$$
\begin{equation}\label{4.20}
\left.
-{\kappa^2 \over 2}\ln {\epsilon^2 \over l^2}
\left(\frac 18 R_{ii}-{1 \over 12} R-m^2\right)
\right\}~~~.
\end{equation}
Here $P$ is the quantity defined after
Eq. (\ref{3.2})
and $l$ is an additional infrared cutoff
parameter imposed
at a large distance from the horizon.
For massive fields $l\simeq m^{-1}$.

The following remarks concerning expressions
(\ref{4.19}) and (\ref{4.20}) are in order.
As compared to the computations
on the Rindler space,
Eqs. (\ref{4.5}),
there are logarithmic corrections from
the mass of the fields and non-zero curvature
near the horizon.
Because of the logarithmic terms
the densities ${dn_i \over d\omega}$
have a different behavior at small frequencies
for scalar and spinor fields.
Spinor density (\ref{4.20}) is
positive at small $\epsilon$ in the whole range
of frequencies. Contrary to this, scalar density
(\ref{4.19}) at $\omega\rightarrow 0$
is proportional
to $\kappa^2m^2\ln \epsilon^2/ l^2$
and is negative (note that curvature corrections
are negligible with respect to the mass of the field).
Such a feature indicates that for scalars
the description of the modes with low frequencies,
the so called soft modes, may need a modification
Ref. \cite{FF:97}.

\bigskip

{}From Eqs. (\ref{1.3}),(\ref{4.19}),(\ref{4.20})
we get the divergent parts of the
canonical free energies
\begin{equation}\label{4.21}
F^C_{s,\mbox{div}}[g,\beta,\epsilon]=
-{1 \over \kappa^3} \int_{\Sigma}
\left\{{\pi^2 \over 180  \beta^4\epsilon^2}
-
\left[{\pi^2 \over 720 \beta^4}P
+
{\kappa^2 \over 48 \beta^2}
\left(\left(\frac 16-\xi\right) R-m^2\right)\right]
\ln {\epsilon^2 \over l^2}
\right\}~~~,
\end{equation}
$$
F^C_{d,\mbox{div}}[g,\beta,\epsilon]=
-{r_d \over \kappa^3}\int_{\Sigma}
\left\{\left({7\pi^2 \over 1440 \beta^4 }+
{\kappa^2 \over 192\beta^2}
\right){1 \over \epsilon^2}\right.
$$
\begin{equation}\label{4.22}
\left.
-\left[{7\pi^2 \over 5760 \beta^4 } P
+{\kappa^2 \over 96 \beta^2}
\left(\frac 18 R_{ii}-{1 \over 12} R-m^2\right)\right]
\ln {\epsilon^2 \over l^2}
\right\}~~~.
\end{equation}
These  equations enable one to calculate
the divergences of the other
characteristics of canonical ensembles. In particular,
one can see that the
statistical-mechanical entropy
diverges near the Killing horizon
and in the leading asymptotic it is
proportional to the area ${\cal A}$
of the horizon. For scalar fields this leading
asymptotic
coincides
with the WKB results by t'Hooft \cite{Hooft:85}
and many other authors.
Equations (\ref{4.21})
and (\ref{4.22}) also follow from the high-temperature
expansions obtained by Dowker and Schofield
\cite{DoSc:88},\cite{DoSc:89}. Application of these
results to our case
is justified because when approaching the horizon
the local temperature unlimitedly grows.

\section{Canonical formulation:
ultraviolet divergences}
\setcounter{equation}0

Comparison of Eq. (\ref{3.5}) with
Eqs. (\ref{4.21}),(\ref{4.22}) shows that
the divergences
of the Euclidean and canonical free energies
are expressed in terms of the similar
geometrical quantities
but have different dependence on the temperature.
Also the nature of the divergences is different:
$F^E$ diverges in an ultraviolet way
while the
$\epsilon$-divergence in $F^C$ has an
infrared origin.
Finally, there is one more important
difference between
the regularizations of $F^E$ and $F^C$.
The ultraviolet regularizations are usually applied
to operators and functionals but not to the background
field itself.
Contrary to this, the volume cutoff regularization
makes the space incomplete and modifies
the background field essentially.

In the presence of the horizon the
densities of levels
have a remarkable property. Namely, there
are regularizations of
${dn \over d\omega}$ which make it possible to
remove the infrared
cutoff near the horizon and to define the densities
on the complete background.
As a result, ${dn \over d\omega}$ acquire
new divergences which correspond exactly to the
ultraviolet divergences of the covariant
Euclidean theory.

\bigskip

As the  first example, let us consider the dimensional
regularization.
The power of the leading divergency in
Eqs. (\ref{4.19}),(\ref{4.20}) is determined
by the dimensionality of the space
$\bar{\cal B}$.
In $D$-dimensional
space-time the leading divergence is $\epsilon^{2-D}$,
if $D\neq 2$, and at $D>2$ one can
take the limit $\epsilon\rightarrow 0$.
After the analytical continuation to the complex values
of $D$ the quantities ${dn \over d\omega}$,
have a pole at $D=4$.
The method how to investigate ${dn \over d\omega}$
near the pole is the following. As before we use
relation (\ref{4.2}).
By taking into account the form  of the
operators $\bar{H}_i^2$, see Eq. (\ref{2.19}),
we can write
\begin{equation}\label{5.1}
\left[ e^{-\bar{H}_i^2t} \right]_{\mbox{diag}}
\simeq
{1 \over (4\pi t)^{(D-1)/2}}e^{-m^2\kappa^2\rho^2 t}
\left(r_i+\bar{b}_{i,1}t+
\bar{b}_{i,2}t^2+...\right)~~~,
\end{equation}
\begin{equation}\label{5.2}
\bar{b}_{s,1}=\kappa^2\rho^2
\left(\frac 16-\xi\right) R
+O(\rho^4)+O(D-4)~~~,
\end{equation}
\begin{equation}\label{5.3}
\bar{b}_{d,1}=r_d {\kappa^2 \over 2}
\left[1+\rho^2\left(\frac 12 R_{ii}-
\frac 12 R_{ijij}-
\frac 16 R\right)+
O(\rho^4)+O(D-4)\right]~~~,
\end{equation}
\begin{equation}\label{5.4}
\bar{b}_{s,2}=O(\rho^4)+O(D-4,\rho^2)~~~,~~~
\bar{b}_{d,2}=\frac 12 r_d \kappa^4
m^2\rho^2+O(\rho^4)+O(D-4,\rho^2)~~~.
\end{equation}
The coefficients $\bar{b}_{i,n}$ are found with the
help of
Eqs. (\ref{4.6}),(\ref{4.14})-(\ref{4.16})
The terms $O(D-4)$, $O(D-4,\rho^2)$
denote additions which appear
in formulas (\ref{4.14})-(\ref{4.16}) when $D\neq 4$.
We do not write these terms explicitly because
they do not result in singularities when
$D\rightarrow 4$.
The integration measure in the traces is
obtained from Eqs. (\ref{4.8})-(\ref{4.11})
\begin{equation}\label{5.5}
\int_{\bar{B}}\sqrt{\bar{g}}d^{D-1}x\simeq
{1 \over \kappa^{D-1}}
\int_{\Sigma}\int_{0}^{\infty}
\rho^{1-D}d\rho\left[1+\frac 14
\rho^2 P+O(D-4,\rho^2)
\right]~~~.
\end{equation}
After substitution of Eqs.
(\ref{5.2})-(\ref{5.4}) in
Eq. (\ref{5.1}) and making use of (\ref{5.5}) we can
derive the singular part of the traces
\begin{equation}\label{5.6}
\left[\mbox{Tr}
e^{-\bar{H}^2_st}\right]_{\mbox{div}}=
{\Gamma\left(1-\frac D2\right) \over (4\pi)^{(D-1)/2}}
 {m^{D-4} \over 2\kappa t^{3/2}} \int_{\Sigma}
\left[\left(m^2-\left(\frac 16 -\xi\right)R\right)t
-{P \over 4\kappa^2}\right]~~~,
\end{equation}
\begin{equation}\label{5.7}
\left[\mbox{Tr}e^{-\bar{H}^2_d t}
\right]_{\mbox{div}}=
r_d{\Gamma\left(1-\frac D2\right)
\over (4\pi)^{(D-1)/2}}
{m^{D-4} \over 2\kappa t^{3/2}} \int_{\Sigma}
\left[\left(m^2+{R\over 12} -
{R_{ii} \over 8}\right)t
-{P \over 4\kappa^2}\right]~~~.
\end{equation}
The $\Gamma$-functions appear as a result
of the integration over $\rho$.
It should be noted that for the spinors the
contribution
of the coefficient $\bar{b}_{d,2}$ cancels the pole
caused by the term $r_d\kappa^2/2$ in $\bar{b}_{d,1}$.
The divergent part of
density of energy levels is obtained from
(\ref{5.6}),(\ref{5.7}) with the help
of Eq. (\ref{4.2})
\begin{equation}\label{5.8}
\left[{dn_s(\omega,D) \over d\omega}
\right]_{\mbox{div}}=
{\Gamma\left(1-\frac D2\right) \over (4\pi)^{D/2}}
 {m^{D-4} \over \kappa} \int_{\Sigma}
\left[2\left(m^2-\left(\frac 16 -\xi\right)R\right)
-{\omega^2 \over \kappa^2} P\right]~~~,
\end{equation}
\begin{equation}\label{5.9}
\left[{dn_d(\omega,D) \over d\omega}
\right]_{\mbox{div}}=
r_d{\Gamma\left(1-\frac D2\right) \over (4\pi)^{D/2}}
 {m^{D-4} \over \kappa} \int_{\Sigma}
\left[2\left(m^2+{R\over 12} -{R_{ii} \over 8}\right)
-{\omega^2 \over \kappa^2}P\right]~~~.
\end{equation}
The divergence represents
a simple pole at $D=4$.
Finally, from Eqs. (\ref{1.3}),(\ref{5.8}),(\ref{5.9})
one
can find the divergent part of
the canonical free energy.
We do not write this divergence explicitly because
it is exactly the same as that of the
Euclidean free energy, Eq. (\ref{3.2}),
computed in the dimensional
regularization
\begin{equation}\label{5.10a}
F^C_{\mbox{div}}[g,\beta,D]=
F^E_{\mbox{div}}[g,\beta,D]~~~.
\end{equation}
This key equality can  be also
established in the Pauli-Villars regularization.
The regularized density of states in this method
is the following quantity
\begin{equation}\label{5.11}
{dn_i(\omega,\mu) \over d\omega}\equiv
{dn_i(\omega) \over d\omega}+\sum_k
{dn_i(\omega,M_k) \over d\omega}
-\sum_r {dn_i(\omega,M'_r) \over d\omega}~~~.
\end{equation}
Definition (\ref{5.11}) takes into account
that in the Pauli-Villars method each field
is replaced by the "multiplet" of fields.
The quantities
${dn_i(\omega,M_k) \over d\omega}$,
${dn_i(\omega,M_r') \over d\omega}$
are the densities of
levels of the Pauli-Villars partners and the fields
with the wrong statistics give negative
contributions.
The number of such fields equals
the number of the fields with the correct statistics
and the leading
$\epsilon$-divergences in Eqs.
(\ref{4.19}),(\ref{4.20})
are cancelled. Regarding logarithmical
divergences $\ln\epsilon^2$, they disappear
because of constraint (\ref{3.4})
with $p=2$. As a result, regularized
densities (\ref{5.11}) are left finite
when $\epsilon\rightarrow 0$ and
can be defined on the complete space.
When the Pauli-Villars cutoff is removed
($\mu\rightarrow \infty$),
${dn_i(\omega,\mu) \over d\omega}$ diverge in
an ultraviolet way. The divergences
can be inferred from Eqs. (\ref{5.8}),(\ref{5.9})
by taking into account constraints (\ref{3.4})
\begin{equation}\label{5.12}
\left[{dn_s(\omega,\mu) \over d\omega}
\right]_{\mbox{div}}=
{1 \over (4\pi)^2 \kappa}
 \int_{\Sigma}
\left[2b+a\left({\omega^2 \over \kappa^2}P+
2\left(\frac 16 -\xi\right)R\right)\right]~~~,
\end{equation}
\begin{equation}\label{5.13}
\left[{dn_d(\omega,\mu) \over d\omega}
\right]_{\mbox{div}}=
r_d {1 \over (4\pi)^2 \kappa}
\int_{\Sigma}
\left[2b+a \left({\omega^2 \over \kappa^2}P
-{R\over 6} +{R_{ii} \over 4}\right)
\right]~~~.
\end{equation}
Constants $a$ and $b$ are defined by
Eqs. (\ref{3.5b}),(\ref{3.5c}) and diverge in the
limit of infinite $\mu$. With the help of Eqs.
(\ref{5.12}),(\ref{5.13}) one can show that
\begin{equation}\label{5.10b}
F^C_{\mbox{div}}[g,\beta,\mu]=
F^E_{\mbox{div}}[g,\beta,\mu]~~~,
\end{equation}
where $F^E_{\mbox{div}}[g,\beta,\mu]$ is given by
Eq. (\ref{3.5}).
It should be noted in conclusion that
Pauli-Villars regularization of the
canonical free energy was first suggested
by Demers, Lafrance and Myers
\cite{DLM:95} who considered a scalar field
on the Reissner-Nordstr\"om black hole background.
The authors used the WKB method.
Although
our method is different the results for
$F^C_{\mbox{div}}[g,\beta,\mu]$ in this particular
case coincide.

\section{Discussion}
\setcounter{equation}0

We are interested in finding the
relation between the covariant Euclidean and
the canonical formulations of statistical
mechanics on curved backgrounds with
horizons.
Let us discuss first why these formulations
are equivalent for spaces without horizons.
As we showed in Section 2, the canonical
formulation
is equivalent to the Euclidean theory
on ultrastatic
background, $\bar{{\cal M}}_\beta$,
Eq. (\ref{2.17}), conformally related to
the original
space-time ${\cal M}_\beta$, Eq. (\ref{2.5}).
The Euclidean actions $W_i$ are determined by the
operators $L_i$ on ${\cal M}_\beta$, see
Eq. (\ref{2.6}).
Analogously, in canonical theory the
functionals
$\bar{W}_i$ are determined by operators
$\bar{L}_i$ on
$\bar{{\cal M}}_\beta$. The classical actions
corresponding to
these two types of operators are
\begin{equation}\label{6.1}
I_i[g,\varphi_i]=\int_{{\cal M}_\beta}
\varphi^{+}_i L_i\varphi_i\sqrt{g}d^4x~~~,~~~
\bar{I}_i[\bar{g},\bar{\varphi}_i]=
\int_{\bar{{\cal M}}_\beta}
\bar{\varphi}^{+}_i \bar{L}_i\bar{\varphi}_i
\sqrt{\bar{g}}d^4x~~~,
\end{equation}
where the notation $\varphi_i$ is used
for scalars $\phi$
or spinors $\psi$. As a result of Eq. (\ref{2.13}),
\begin{equation}\label{6.1a}
I_i[g,\varphi_i]=\bar{I}_i[\bar{g},\bar{\varphi}_i]
\end{equation}
for $\bar{\phi}=e^{-\sigma}\phi$
and $\bar{\psi}=e^{-\frac 32 \sigma}\psi$.
The transformation
from one action to another is not singular
and the classical theories on  ${\cal M}_\beta$
and $\bar{{\cal M}}_\beta$ are equivalent.
In case of massless spinors and massless scalars
with $\xi=\frac 16$ the operators
$L_i$ and $\bar{L}_i$
have the same form, which means that the
classical theories
are conformally invariant.
In general case this invariance does not exist.
However, it is still possible to
introduce an auxiliary
conformal charge in the classical actions and
interpret Eq. (\ref{6.1a}) in terms of a
pseudo conformal invariance
\cite{DoSc:88},\cite{DoSc:89}.
According to a common point of view
\cite{Duff:77},\cite{BiDa:82} the
{\it bare} quantum actions respect the
classical symmetries. Thus, for the
bare regularized functionals there is the same
equality as for the classical
actions\footnote{It is
true if the regularization itself does not break
classical symmetries},
\begin{equation}\label{6.2}
W_i[g,\beta]_{\mbox{bare}}=
\bar{W}_i[g,\beta]_{\mbox{bare}}~~~.
\end{equation}
This relation is not valid for the
{\it renormalized} quantities because
the conformal symmetry
is broken by the quantum
anomalies \cite{Duff:77},\cite{BiDa:82}.
The difference
between the renormalized actions $W_i$ and
$\bar{W}_i$ for scalar and spinor
fields was found explicitly by Dowker and Schofield
\cite{DoSc:88},\cite{DoSc:89}. The anomaly,
which is an integral over the Euclidean space,
is proportional to $\beta$ and so it contributes
to the vacuum energy only. As a result,
the free energies $F_i^E$ and $F_i^C$,
Eqs. (\ref{2.8}),(\ref{2.10}), coincide
{\it before} and {\it after}
renormalization.

\bigskip

In case of the horizons
${\cal M}_\beta$ and $\bar{{\cal M}}_\beta$
have the different topologies,
$\R^2\times \Sigma$ and
$S^1\times \bar{\cal B}$, respectively.
The transformation which relates
${\cal M}_\beta$ and $\bar{{\cal M}}_\beta$ is
singular on the
bifurcation surface $\Sigma$.
So
the classical theories are not quite equivalent.
On the quantum level the horizons result
in the divergences of $F_i^E$ and $F_i^C$ which
cannot be eliminated by subtracting the vacuum
energy. Moreover, the divergences
of $F_i^E$ and $F_i^C$
have the different origins.

Our results suggest a way
how can  the covariant Euclidean and
the canonical formulations
be reconciled.
We showed that there are regularizations which are
applicable to both  $F^E_i$
and  $F^C_i$.
In such regularizations the canonical
free energy $F^C_i$
can be defined on the complete background
$\bar{{\cal M}}_\beta$ and its divergences
are identical to the divergences
of $F^E_i$,
see Eqs. (\ref{5.10a}),(\ref{5.10b}).
When the functionals $F^E_i$ and $F^C_i$
are well defined,
it becomes possible to carry out the transformation
from ${\cal M}_\beta$ to $\bar{{\cal M}}_\beta$
and to interpret it as a conformal symmetry.
In analogy with the case without horizons, we can
make a hypothesis
that (at least for scalars and spinors)
the regularized
{\it bare} free energies
are identical
\begin{equation}\label{6.3}
F_i^E[g,\beta]_{\mbox{bare}}=
F_i^C[g,\beta]_{\mbox{bare}}~~~.
\end{equation}
It is assumed that the both
functionals in (\ref{6.3})
are considered in the same regularization.
To give a strict proof of this equality
may be a
rather difficult problem.
There are examples which enable one
its direct check. In Ref. \cite{CVZ:95}
Cognola, Vanzo and Zerbini obtained
the free energy of massive
scalar fields in the Rindler space in an
explicit form,
see also
Ref. \cite{BCZ:96}. It can be shown that
these results,
rewritten in the Pauli-Villars
regularization, confirm relation (\ref{6.3}).
Another direct
confirmation is possible in two dimensions.
Two dimensional massless scalar fields were
analysed in Ref. \cite{Solodukhin:96}
and these results support our hypothesis as well.

Equality (\ref{6.3}) enables one to apply
the methods of quantum field theory
to statistical mechanics with the horizons.
As a consequence, it justifies the
ultraviolet renormalization of
statistical-mechanical quantities \cite{FS:96}.
Our results
are also important for studying
the statistical-mechanical
foundation of the thermodynamics of
black holes. It was realized in the last
years that statistical-mechanical computations
in this case require an off-shell
procedure \cite{FFZ:96a},
i.e., considering a black hole
at temperatures different from the Hawking value.
{}From this point of view,
Eq. (\ref{6.3}) demonstrates
the equivalence of two off-shell methods,
the canonical and  the conical
singularity methods.
However, the comparison of the
off-shell and on-shell
results goes beyond the subject
of this paper.
As non-minimally coupled scalars
show \cite{FF:97},
the off-shell and on-shell computations
are not always equivalent.

Finally, several remarks about the range
of validity of our
results are in order.
We dealt with static nonextremal black hole
backgrounds.
The method described in Sections 4, 5 is
applicable to the extremal
black holes as well, but rotating black holes
require an additional analysis.
For the extremal black
holes the density of levels
${d n\over d\omega}$ has the same property
as in nonextremal case. Namely,
the Pauli-Villars and
dimensional regularizations eliminate the
$\epsilon$-divergences.
For the scalar fields the corresponding calculations
were done in Ref. \cite{DLM:95}.
Our consideration was also restricted to
the scalar and spinor fields.  The method to calculate
the quantity ${d n\over d\omega}$
can be generalized to the fields of other
spins.
Finding the correspondence between the canonical and
the Euclidean formulations for these fields
would be an interesting extension of
this work.

\vspace{12pt}
{\bf Acknowledgements}:\ \ I am grateful to
V. Frolov, Yu. Gusev,
and A. Zelnikov for
helpful discussions.
This work was
partially supported  by the Natural
Sciences and Engineering
Research Council of Canada.

\newpage
\appendix
\section{Canonical free-energy and effective action}
\setcounter{equation}0

Here we give the details how to obtain
relation (\ref{2.10}).
We suppose that the system has
a discrete spectrum of frequencies.
Then the canonical free
energy is
\begin{equation}\label{ap1.0}
F^C_i[g,\beta]=\eta_i\beta^{-1}
\sum_{\omega} d_i(\omega)
\ln \left(1-\eta_i e^{-\beta \omega}\right)~~~,
\end{equation}
where  $\eta_s=1$, $\eta_d=-1$ and
$d_i(\omega)$
is the degeneracy of the level $\omega$.
Equation (\ref{1.3}) for $F^C_i[g,\beta]$
is obtained
in the limit when intervals between the
frequencies $\omega$
go to zero.
The basic identities we use are \cite{GrRy:94}
\begin{equation}\label{ap1.3}
\ln \left(1-e^{-\beta \omega}\right)=
-{\beta \omega \over 2}
+\ln \beta \omega +
\sum_{k=1}^{\infty}\ln\left(1+\omega^2
{\beta^2 \over
4 k^2\pi^2}
\right)~~~,
\end{equation}
\begin{equation}\label{ap1.4}
\ln \left(1+e^{-\beta \omega}\right)=
-{\beta \omega \over 2}+\ln 2
+ \sum_{k=0}^{\infty}\ln\left(1+\omega^2 {\beta^2
\over (2k+1)^2\pi^2} \right)~~~.
\end{equation}
Note that
$$
\sum_{k=1-q}^{\infty}\ln\left(1+
\omega^2 {\beta^2 \over
(2 k+q)^2\pi^2}
\right)
$$
\begin{equation}\label{ap1.5}
=
-\lim_{z\rightarrow 0}{d \over dz}
\left[\sum_{k=1-q}^{\infty}
\left({\pi^2 (2k+q)^2 \over \beta^2}
+\omega^2\right)^{-z}-
\sum_{k=1-q}^{\infty}
\left({\pi^2 (2k+q)^2 \over \beta^2}
\right)^{-z}\right]
\end{equation}
where $q=0$ or $1$. By using the properties of
the Riemann $\zeta$ function
\cite{GrRy:94} we find
\begin{equation}\label{ap1.6}
\lim_{z\rightarrow 0}{d \over dz}
\sum_{k=1}^{\infty}
\left({\pi^2 (2k)^2 \over \beta^2}
\right)^{-z}=-\ln \beta~~,~~
\lim_{z\rightarrow 0}{d \over dz}
\sum_{k=0}^{\infty}
\left({\pi^2 (2k+1)^2 \over \beta^2}
\right)^{-z}=-\ln 2~~.
\end{equation}
Thus, Eqs.  (\ref{ap1.3}) and (\ref{ap1.4}) are
represented in the form
\begin{equation}\label{ap1.7}
\ln \left(1-\eta_i
e^{-\beta \omega}\right)=-{\beta \omega \over 2}
-\frac 12 \lim_{z\rightarrow 0}{d \over dz}
\zeta_i(z~|\omega,\beta)~~~,
\end{equation}
\begin{equation}\label{ap1.8}
\zeta_i(z~|\omega,\beta)=
\sum_{k=-\infty}^{\infty}\left[\left(
{2\pi \over \beta}(k+l_i)\right)^2
+\omega^2\right]^{-z}~~~,
\end{equation}
where $l_s=0$ and $l_d=\frac 12$.
The series (\ref{ap1.8}) converge at
${\mbox Re} z >\frac 12$, so the functions
$\zeta_i(z~|\omega,\beta)$ can be defined at
$z\rightarrow 0$ with the help of
the analytic continuation.
By taking into account Eqs. (\ref{ap1.0}) and
(\ref{ap1.7}) we get the canonical
free energy in form (\ref{2.10})
\begin{equation}\label{ap1.9}
F^C_i[g,\beta]=\beta^{-1}\bar{W}_i[g,\beta]
-\bar{E}^0_i[g]~~~,
\end{equation}
\begin{equation}\label{ap1.9a}
\bar{W}_i[g,\beta]=\eta_i{1 \over 2}
\sum_{\omega} d_i(\omega)
\lim_{z\rightarrow 0}{d \over dz}
\zeta_i(z~|\omega,\beta)~~~,~~~
\bar{E}^0_i[g]=\eta_i\sum_{\omega}
d_i(\omega){\omega \over 2}~~~,
\end{equation}
where $\bar{E}^0_i[g]$ is the vacuum energy.
The quantities $\bar{W}_i[g,\beta]$ and
$\bar{E}^0_i[g]$ diverge at large
$\omega$ although
the free energy itself is finite.
To see that
$\bar{W}_i[g,\beta]$ is effective
action (\ref{2.12})
one can regularize this functional
with the help of the $\zeta$ function method.
It is enough to make the replacement
\begin{equation}\label{ap1.10}
\sum_{\omega}d_i(\omega)
\lim_{z\rightarrow 0}{d \over dz}
\zeta_i(z|\omega,\beta)\rightarrow
\lim_{z\rightarrow 0}
{d \over dz}\zeta_i(z|\beta)
\end{equation}
where , according to Eq. (\ref{ap1.8}),
\begin{equation}\label{ap1.11}
\zeta_i(z|\beta)=\sum_{\omega}
\sum_{k=-\infty}^{\infty}
d_i(\omega)\left[\left(
{2\pi \over \beta}(k+l_i)\right)^2
+\omega^2\right]^{-z}~~~.
\end{equation}
$\zeta_i(z|\beta)$ are the
generalized $\zeta$ functions of the operators
$\bar{L}_i$, see Eqs. (\ref{2.14}),(\ref{2.15}).
For operators on manifolds  without boundaries
$\zeta_i(z~|\beta)$ can be defined as a
meromorphic function with
simple poles at $z=1,2$ \cite{Camporesi:90}.

\section{Spinor heat coefficients
on conical singularities}
\setcounter{equation}0

Let us consider the heat kernel of the  spinor
Laplace operator
$\triangle_d=-\nabla^\mu\nabla_\mu +
\frac 14 R$ on spaces ${\cal M}_\beta$
with conical singularities.
For simplicity we put $\kappa=1$, so
${\cal M}_\beta$ are regular when $\beta=2\pi$.
The heat kernel obeys the
equation
\begin{equation}\label{ab.1}
(\triangle_d)_xK_\beta\left(x,x',s\right)+
\partial_sK_\beta\left(x,x',s\right)=0~~~,~~~
K_\beta\left(x,x',0\right)=\delta(x,x')~~~,
\end{equation}
where $\delta(x,x')$ is the delta function on
${\cal M}_\beta$. We first describe the heat kernel
on a simple cone  ${\cal C}_\beta$, which will be
required for us later.
It is known since Sommerfeld
that heat kernels on ${\cal C}_\beta$ can be
expressed
in terms of the corresponding heat kernels
on the plane
$\R^2$. A suitable generalization of the Sommerfeld
representation
for integer and half-odd-integer spins was given
by Dowker \cite{Dowker:77},\cite{Dowker:87}.
By making use of the results of Ref.
\cite{Dowker:87}
we can represent the spinor heat kernel on
${\cal C}_\beta$
in the following form
\begin{equation}\label{ab.2}
K_\beta\left(x(\tau),x'(0),s\right)=
{1 \over 2i\beta}\int_{A}
{1 \over \sin{\pi \over \beta}
(z+\tau)}U(z)K\left(x(z),x'(0),s\right)~~~.
\end{equation}
Here $\tau$ is the polar-angle
coordinate on ${\cal C}_\beta$.  $A$ is the
contour in the complex plane which has two
parts. In the upper half-plane it runs from
$(\pi-\epsilon)+i\infty$ to
$(-\pi+\epsilon)+i\infty$
and in the lower half-plane from
$(-\pi+\epsilon)-i\infty$ to
$(\pi-\epsilon)-i\infty$.
$K\left(x,x',s\right)$ is
the spinor heat kernel
which obeys the problem (\ref{ab.1})
on $\R^2$. The operators
$K_\beta\left(x,x',s\right)$
and $K\left(x,x',s\right)$ correspond
to the different
spin structures and have the
different periodicity.
The kernel on $\R^2$ is unchanged when going
around the origin of the polar coordinate system.
(There is no difference between the origin
of the polar coordinates
and other points on the plane).
Contrary to this, the kernel on
${\cal C}_\beta$ changes the sign
when $\tau$ is increased by $\beta$.
The covariant derivatives on $\R^2$ are trivial,
$\nabla_\mu=\partial_\mu$, but
the
the covariant derivatives on ${\cal C}_\beta$
are defined by the polar tetrades and nontrivial,
$\nabla_\mu=\partial_\mu+
\Gamma_\mu$.
In the basis $\gamma_\mu=(\sigma_1,
\sigma_2)$, where  $\sigma_k$ are the Pauli matrices,
the spinor connection 1-form is
$\Gamma=-\frac i2 \sigma_3 d\tau$.
The role of the matrix $U$ in relation
(\ref{ab.2}) is to make a gauge-like
transformation from the derivative on
$\R^2$ to that on ${\cal C}_\beta$,
\begin{equation}\label{ab.3}
U(\tau)\partial_\mu
U^{-1}(\tau)=\nabla_\mu~~~,~~~
U(\tau)=\exp\left(\frac i2 \sigma_3\tau\right)~~~.
\end{equation}
In fact, when $\tau$ is real $U(\tau)$ is
the unitary matrix which is the
spinor representation
of the rotation on the angle $\tau$.

To find corrections to the heat
coefficients from the
conical singularities we follow
the method suggested in
Ref. \cite{Fursaev:94b}.
According to this method,
it is sufficient to work
in a narrow domain $\tilde{\Sigma}$
of the singular surface $\Sigma$.
The rest region of ${\cal M}_\beta$
does not have conical singularities and
the heat kernel expansion on it has a standard form.
$\tilde{\Sigma}$ can be approximated as
${\cal C}_\beta\times \Sigma$. So here one can
relate $K_\beta(x,x',s)$ with the kernel
$K(x,x',s)$ on the regular space
${\cal M}_{\beta=2\pi}$ by
the formula analogous to
Eq. (\ref{ab.2}).
The contribution
$\mbox{Tr}_{\tilde{\Sigma}}
K_\beta(s)$ from $\tilde{\Sigma}$ to the trace
can be written as
\begin{equation}\label{ab.2b}
\mbox{Tr}_{\tilde{\Sigma}} K_\beta(s)=
\mbox{Tr}_{\tilde{\Sigma}} K(s)+
{1 \over 2i\beta}\int_{A'}
{1 \over \sin{\pi \over \beta}
z}\int_{\tilde{\Sigma}}\sqrt{g} d^Dx
\mbox{Tr}_i \left[U(z)
K\left(x(z),x(0),s\right)\right]~~~,
\end{equation}
where $\mbox{Tr}_i$ stands for the
trace over the spinor indexes.
The points $x(z)$ and $x(0)$ are connected
by the integral
line of the Killing field $\partial_\tau$.
The two terms in r.h.s. of
(\ref{ab.2b}) appear when
contour $A$ is deformed to a
small loop around the origin
and contour $A'$ which consists
of two vertical curves.
The effect of conical
singularities is related to the second term
in (\ref{ab.2b}).
The asymptotic form of $K(x,x',s)$ is
\begin{equation}\label{ab.4}
K(x,x',s)\simeq{e^{-\sigma^2(x,x')/4s}
\over (4\pi s)^{D/2}}
\triangle^{1/2}(x,x')\sum_{n}a_n(x,x')s^n~~~,
\end{equation}
where $\sigma(x,x')$ is the geodesic distance between
points $x,x'$ and $\triangle(x,x')$ is the Van Vleck
determinant. The coefficients $a_n$ are
determined in terms
of the Riemann tensor and its derivatives.
Let $\rho$ be the proper
distance from the points $x(z)$ and
$x(0)$ to $\Sigma$.
One
can find the following relations \cite{Fursaev:94b}
\begin{equation}\label{ab.5}
\sigma^2(x(z),x(0))\simeq 4\rho^2\sin^2
\frac z2-\frac 16\rho^4R_{ijij}~ \sin^2z
{}~~~,
\end{equation}
\begin{equation}\label{ab.6}
\triangle^{1/2}(x(z),x(0))\simeq 1+\frac 16 \rho^2
R_{ii} ~\sin^2\frac z2~~~,
\end{equation}
where $R_{ii}$ and $R_{ijij}$ are
defined in (\ref{3.3}).
The integration measure on $\tilde{\Sigma}$
can be derived from Eqs. (\ref{4.8})-(\ref{4.10})
\begin{equation}\label{ab.7}
\int_{\tilde{\Sigma}} \sqrt{g} d^Dx
\simeq \int_{\Sigma}
\sqrt{\gamma}d^{D-2}\theta\int \rho d\rho d\tau
\left[1+\rho^2\left(\frac 16 R_{ijij}-
\frac 14 R_{ii}\right)\right]~~~.
\end{equation}
For the first coefficients in Eq. (\ref{ab.4})
one finds
\begin{equation}\label{ab.8}
a_0(x',x)\simeq I+\frac 18 R_{\mu\alpha\lambda\nu}
\Sigma^{\nu\lambda}(x')^\mu x^\alpha~~~,
{}~~~a_1(x',x)\simeq -{1 \over 12} R I~~~,
\end{equation}
where $I$ is the unit matrix in the
spinor representation
and $\Sigma^{\mu\nu}=
\frac 14[\gamma^\mu,\gamma^\nu]$.
The matrix $U(z)$ corresponds to the rotation
of a vector normal to the surface $\Sigma$ on
the angle $z$. So from Eq. (\ref{ab.8}) we get
\begin{equation}\label{ab.9}
\mbox{Tr}_i\left[U(z)a_0
\left(x(z),x(0)\right)\right]
\simeq r_d\cos \frac z2
\left(1+\frac 14 \rho^2\sin^2
\frac z2 R_{ijij}\right)~~~.
\end{equation}
Here $r_d$ is the dimensionality of the
spinor representation. Now, it follows
from
Eqs. (\ref{ab.4})-(\ref{ab.6}),(\ref{ab.8}) that
$$
\mbox{Tr}_i\left[U(z)K_\beta(x(z),x(0))\right]
$$
\begin{equation}\label{ab.10}
\simeq r_d{\exp\left({-{\rho^2 \over s}
\sin^2\frac z2}\right)
 \over (4\pi s)^{D/2}}\cos \frac z2 \left[1
+\rho^2\left(\frac 14 R_{ijij}+
\frac 16 R_{ii}\right) \sin^2\frac z2 +
{\rho^4 \over 24s}R_{ijij}\sin^2z-
{1 \over 12} Rs\right]~~~.
\end{equation}
It can be shown that approximation (\ref{ab.10})
is sufficient to find
all corrections for the first three heat
coefficients due to conical
singularities.
By integrating (\ref{ab.10})
with measure (\ref{ab.7}) we obtain an integral
over the surface $\Sigma$. Its integrand is an
expression  linear in the quantities $R$, $R_{ii}$
and $R_{ijij}$ with coefficients proportional
to $\cos \frac z2
\sin^{-2q} \frac z2$, $q=1,2$. Finally, we have to
integrate this expression in the complex
plane, see. Eq. (\ref{ab.2b}). This can be done
with the help of formulas  \cite{Dowker:94b}
\begin{equation}\label{ab.11}
{1 \over i\beta} \int_{A'}
{dz \over \sin {\pi \over \beta} z}~ {\cos \frac z2
\over \sin^2 \frac z2}=-{ 1 \over 3} (\gamma^2-1)~~~,
\end{equation}
\begin{equation}\label{ab.12}
{1 \over i\beta} \int_{A'}
{dz \over \sin {\pi \over \beta} z}~ {\cos \frac z2
\over \sin^4 \frac z2}=-{1 \over 180}(\gamma^2-1)
(7\gamma^2+17)~~~,
\end{equation}
where $\gamma={2\pi \over \beta}$ (or, in general
case, $\gamma={\beta_H \over \beta}$). By making
use of
these relations and Eq. (\ref{ab.2b}) we find
\begin{equation}\label{ab.13}
\mbox{Tr}_{\tilde{\Sigma}} K_\beta(s)-
\mbox{Tr}_{\tilde{\Sigma}} K(s)\simeq
{1 \over (4\pi s)^{D/2-1}}\left(A_{\beta,1}+s
A_{\beta,2}+O(s^2)\right)~~~,
\end{equation}
where $A_{\beta,k}$ are given
by Eqs. (\ref{3.10}),(\ref{3.11}) and Table 1
for spinors.
Formula (\ref{ab.13}) determines the difference
between asymptotic expansions on regular and
singular spaces. It is easy to see
that $A_{\beta,k}$
are the corrections to the spinor heat
coefficients from conical
singularities.

\end{document}